\documentclass[twocolumn,prb,amssymb,showpacs]{revtex4}
\usepackage{graphicx}
\usepackage{dcolumn}
\usepackage{amsmath}

\usepackage{graphicx}
\begin{document}

\title{Antilocalization and spin-orbit coupling in hole strained
GaAs/In$_x$Ga$_{1-x}$As/GaAs quantum well heterostructures}

\author{G.~M.~Minkov}
\author{A.~A.~Sherstobitov}
\affiliation{Institute of Metal Physics RAS, 620219 Ekaterinburg,
Russia}

\author{A.~V.~Germanenko}
\author{O.~E.~Rut}
\author{V.~A.~Larionova}
\affiliation{Institute of Physics and Applied Mathematics, Ural
State University, 620083 Ekaterinburg, Russia}

\author{B.~N.~Zvonkov}

\affiliation{ Physical-Technical Research Institute, University of
Nizhni Novgorod,  603600 Nizhni Novgorod, Russia}

\begin{abstract}
{Low-field  magnetoresistance in $p$-type strained quantum wells
is studied. It is shown that the Rashba mechanism leads to the
cubic in quasimomentum spin-orbit splitting of the hole energy
spectrum and the antilocalization behavior of low-field
magnetoresistance is well described by the Hikami-Larkin-Nagaoka
expression. }
\end{abstract}
\pacs{73.20.Fz, 73.61.Ey}

\maketitle

The combination of quantum coherence and spin rotation produces a
number of interesting transport properties. Numerous proposals for
electronic devices  that use spin-orbit coupling have appeared in
last years, including gate-controlled sources and detectors of
spin-polarized current. \cite{spdev1,spdev2,spdev3} Spin-orbit
coupling results in the spin splitting of the energy spectrum when
an inversion symmetry is lifted. The lack of inversion symmetry of
the original crystal results in the splitting of the energy
spectrum, which is linear and cubic in in-plane quasimomentum,
$k$. This splitting is described by terms known as the Dresselhaus
terms.\cite{Mindress} In low-dimensional systems an additional
mechanism of spin splitting is caused by the asymmetry of the
confining potential (so called the Rashba term\cite{Minrashba}).
In two-dimensional (2D) semiconductor systems this asymmetry
arises from asymmetry of the smooth electrostatic potential in the
perpendicular to the 2D plane direction, from Schottky barrier
potential, from asymmetry in doping layers disposition, and
composition gradient along the growth direction. It is very
important that this asymmetry can be controlled by gate voltage.
For electron 2D states, the Rashba term  is linear  in $k$. For
hole 2D systems, the situation becomes more complicated because of
four-fold degeneracy of the topmost valence band $\Gamma_8$ of the
parent material. Theoretical considerations of this problem and
experimental studies show that the splitting is cubic in $k$ in
this case.\cite{Gerchikov,Winkler,Winkler1,Xu}

The measurements of interference induced low-filed
magnetoresistance are the powerful  tool for studies of the
spin-splitting, spin- and phase- relaxation mechanisms. At
present, there are numerous studies of $n$-type 2D
systems\cite{chen93,spdev1,Knap96,Knap96pss,Zdun,Hass97,Miller03,Studen03},
whereas the more complicated $p$-type systems are studied
noticeable less\cite{Peder99,Schier02,Parad02,Golub02,Proskur02}
(for more references see review article by Zawadzki and
Pfeffer\cite{Zawadzki}). As for the strained quantum well, the
antilocalization and spin relaxation in 2D hole gas are
practically not investigated in these systems.

In this paper, we present results of  experimental  study of the
low-field  magnetoresistance caused by spin relaxation in $p$-type
strained GaAs/In$_x$Ga$_{1-x}$As/GaAs quantum well structures. It
has been found that the magnetoresistance shape is well described
by the Hikami-Larkin-Nagaoka (HLN) expression\cite{HLN} that means
that the leading term in the spectrum splitting is cubic in
quasimomentum. We show that in contrast to $n$-type systems, where
such a finding implies that the Dresselhaus spin splitting
mechanism is the main, the Rashba mechanism is responsible for the
spin splitting of the hole energy spectrum in strained quantum
wells under investigation.

\begin{figure}[t]
\includegraphics[width=\linewidth,clip=true]{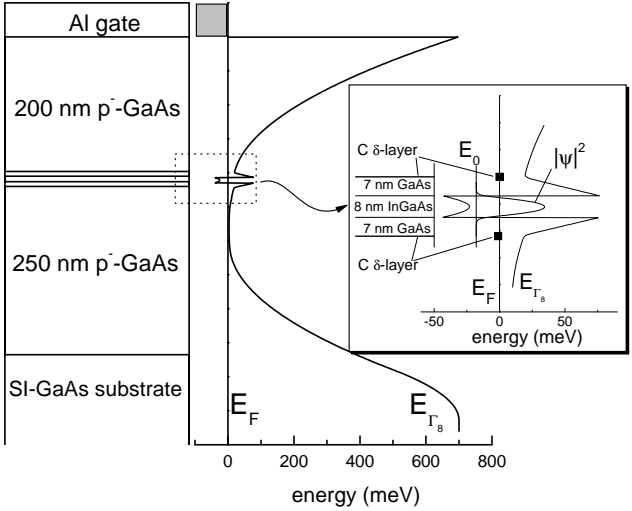}
\caption{\label{f0} The cross section of structure 3857 and its
energy diagram. }
\end{figure}

The GaAs/In$_x$Ga$_{1-x}$As/GaAs heterostructures were grown by
metal-organic vapor phase epitaxy on semi-insulator  GaAs
substrate. The quantum well was biaxially compressed due to the
lattice mismatch between In$_x$Ga$_{1-x}$As and GaAs. Two types of
heterostructures were studied. The structures of the first type,
3855, 3856, and 3857,  consist of a $250$~nm-thick undoped GaAs
buffer layer, carbon $\delta$-layer, a 7~nm spacer of undoped
GaAs, a 8~nm In$_{0.2}$Ga$_{0.8}$As well, a 7~nm spacer of undoped
GaAs, a carbon $\delta$-layer   and 200~nm cap layer of undoped
GaAs (see Fig.~\ref{f0}). The structure of the second type, 3951,
was analogous, the only difference was the wider spacer, 15~nm,
and as sequence the higher mobility.  The structures within the
first group differ by carbon density in $\delta$-layers. The
parameters of the structures are presented in Table \ref{tab1}.
The samples were mesa etched into standard Hall bars and then an
Al gate electrode was deposited by thermal evaporation onto the
cap layer through a mask. Varying the gate voltage $V_g$ from
$-1$~V to $+3$~V we changed the hole density in the quantum well
from $3\times 10^{11}$~cm$^{-2}$ to $1\times 10^{12}$ cm$^{-2}$.
The analysis of the temperature dependence of the Shubnikov-de
Haas oscillations showed that the hole effective mass was equal to
$(0.160\pm 0.005)$\,$m_0$ and did not depend on the hole density.

\begin{table}[b]
\caption{The parameters of structures investigated} \label{tab1}
\begin{ruledtabular}
\begin{tabular}{crccc}
Structure & $N_1$ (cm$^{-2}$)\footnotemark[1] & $N_2$
(cm$^{-2}$)\footnotemark[1] & $p$
(cm$^{-2}$) & $\mu$ (cm$^2$/V\,s)  \\
 \colrule
 3855          &$4\times10^{11}$   &$3\times10^{11}$   &$4.7\times 10^{11}$ & 4800 \\
 3856          &$8\times10^{11}$  &$6\times10^{11}$&$7.5\times 10^{11}$ & 5700\\
 3857          &$1.2\times10^{12}$  &$8\times10^{11}$   &$9.5\times 10^{11}$ & 8000\\
 3951          &$1.2\times10^{12}$  &$8\times10^{11}$ &$5.4\times 10^{11}$& 13100\\
\end{tabular}
\end{ruledtabular}
 \footnotetext[1]{$N_1$ and $N_2$ are the carbon density in outer and inner $\delta$-layers, respectively.}
\end{table}

\begin{figure}[t]
\includegraphics[width=\linewidth,clip=true]{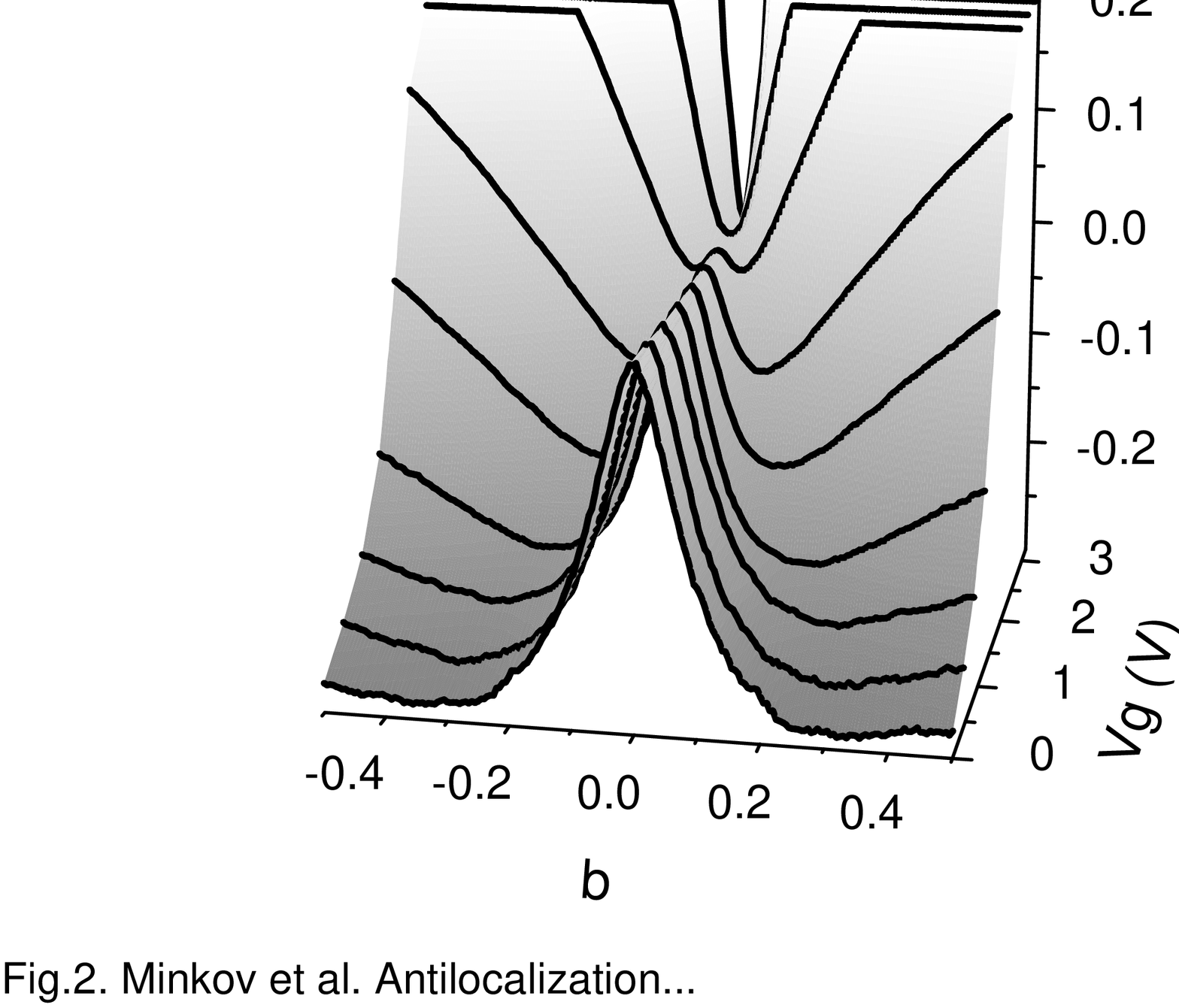}
\caption{\label{f1} The magnetoconductivity plotted against the
reduced magnetic field for different gate voltages,  structure
3857, $T=0.44$~K.}
\end{figure}

Figure~\ref{f1} shows the low-field magnetoconductivity,
$\Delta\sigma(B)=\rho_{xx}^{-1}(B)-\rho_{xx}^{-1}(0)$, measured at
$T=0.44$~K for structure 3857 as a function of a normalized
magnetic field $b=B/B_{tr}$, where $B_{tr}= \hbar/(2e l^2)$ with
$l$ as the mean free path,  is presented in Fig.~\ref{f1}. The
antilocalization maximum at $B=0$ in the
conductivity-versus-magnetic field curves decreases with lowering
hole density and disappears at $V_g=2.75$ V, when $p\simeq
3.8\times 10^{11}$\,cm$^{-2}$.  In the structures 3951 with the
higher hole mobility, the disappearance happens at $p\simeq
3\times 10^{11}$\,cm$^{-2}$.

Theoretically, the low-field anomalous magnetoresistance was
studied in Refs.\,\onlinecite{HLN,Iord,Knap96}. It was shown that
when the spin splitting is cubic in $k$, the magnetoconductivity
curve should be described by the Hikami-Larkin-Nagaoka (HLN)
expression
\begin{eqnarray}
{\Delta\sigma(b)\over G_0}&=&\psi\left({1 \over 2}+{\tau_1\over
b}\left[{1\over\tau_\phi}+{1\over\tau_s}\right]\right)-\ln{\left({\tau_1\over
b}\left[{1\over\tau_\phi}+{1\over\tau_s}\right]\right)} \nonumber\\
&+&{1\over 2}\psi\left({1 \over 2}+{\tau_1\over
b}\left[{1\over\tau_\phi}+{2\over\tau_s}\right]\right)-{1\over
2}\ln{\left({\tau_1\over b}\left[{1\over\tau_\phi}+{2\over\tau_s}\right]\right)}\nonumber\\
&-& {1\over 2}\psi\left({1 \over 2}+{\tau_1\over
b}{1\over\tau_\phi}\right) +{1\over 2}\ln{\left({\tau_1\over
b}{1\over\tau_\phi}\right)}.
 \label{eq1}
\end{eqnarray}
Here, $G_0=e^2/(2\pi^2\hbar)$, $\tau_\phi$ and $\tau_s$ are the
phase and spin relaxation times, respectively, $\psi(x)$ is the
digamma function,  and $\tau_n$, $n=1$, is the transport
relaxation time
\begin{equation}
{1 \over \tau_n}=\int W(\theta)(1-\cos n\theta)d\theta,
\label{eq21}
\end{equation}
where $W(\theta)$ stands for the probability of the scattering by
an angle $\theta$.

For the Dyakonov-Perel spin-relaxation mechanism,\cite{DP} which
is dominant at low temperatures, the value of $\tau_s$ is
determined by the spin-orbit splitting of the energy spectrum,
$\hbar\Omega_3\propto k^3$, as follows
\begin{equation}
{1 \over \tau_s}= 2\Omega_3^2\tau_3 \label{eq2}
\end{equation}
where $\tau_3$ is defined by Eq.~(\ref{eq21}).

Taking into account both the cubic and linear terms leads to more
complicated expression which was obtained in
Ref.\,\onlinecite{Iord}. The following two parameters describing
the spin relaxation arise in this case
\begin{equation}
{1 \over \tau_s'}= 2\Omega_1^2\tau_1 \label{eq3}
\end{equation}
and
\begin{equation}
{1 \over \tau_s}= 2(\Omega_1^2\tau_1+\Omega_3^2\tau_3),\label{eq4}
\end{equation}
where $\hbar\Omega_1$ is the linear in $k$, $\hbar\Omega_1\propto
k$, spin-orbit splitting.

\begin{figure}[t]
\includegraphics[width=\linewidth,clip=true]{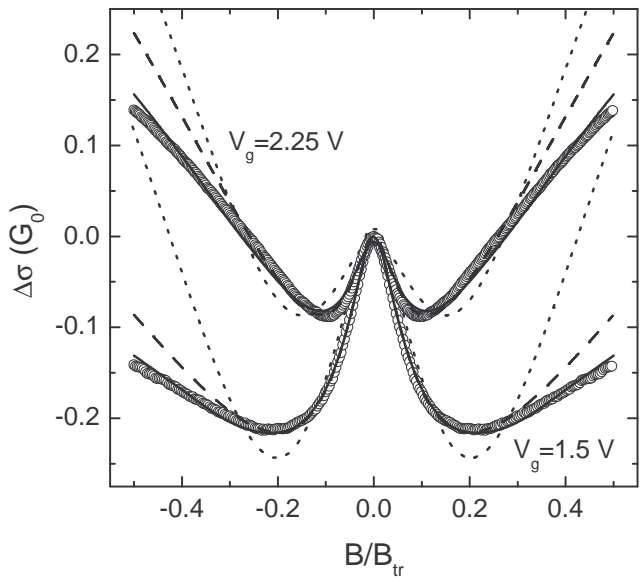}
\caption{\label{f2} The magnetoconductivity as a function of
normalized magnetic field for the two gate voltages: $V_g=1.5$~V
($p=8\times 10^{11}$~cm$^{-2}$, $\tau_1=5.4\times 10^{-13}$~s) and
$V_g=2.25$~V ($p=5.8\times 10^{11}$~cm$^{-2}$, $\tau_1=3.9\times
10^{-13}$~s);  structure 3857, $T=0.44$~K. Symbols are the
experimental data. The dotted lines are the best fit by the
expression from Ref.~\onlinecite{Iord} when only the linear in $k$
term is taken into account. Dashed lines are the best fit by the
HLN expression, Eq.~(\ref{eq1}). Solid lines are the results of
the simulation procedure (see Appendix \ref{app}) which is valid
beyond the diffusion approximation. The fit has been done within
the magnetic field range $-0.3 B_{tr}<B<0.3 B_{tr}$. The fitting
parameters are given in Table~\ref{tab2}.}
\end{figure}

\begin{table}[b]
\caption{The parameters of the best fit for the data presented in
Fig.~\ref{f2} as obtained taking into account only the linear in
$k$ term,\cite{Iord} only the cubic in $k$ term in the diffusion
approximation,\cite{HLN} and the cubic in $k$ term beyond the
diffusion approximation (see Appendix \ref{app}). } \label{tab2}
\begin{ruledtabular}
\begin{tabular}{cccc}
$V_g$ (V) & Theory & $\tau_1/\tau_\phi$ & $\tau_1/\tau_s$
\\
 \colrule
           &Ref.~\onlinecite{Iord}   &0.020   &0.178 \\
 1.5                &Ref.~\onlinecite{HLN}  &0.016   &0.051 \\
                &Appendix \ref{app}   &0.014   &0.040 \\
\colrule
          &Ref.~\onlinecite{Iord}   &0.034   &0.142 \\
2.25                &Ref.~\onlinecite{HLN}  &0.017   &0.032 \\
                &Appendix \ref{app}   &0.013   &0.025
\end{tabular}
\end{ruledtabular}
\end{table}

Comparison of the experimental data with theoretical expressions
for two limiting cases, when only  the cubic or  linear term is
taken into account, is shown in Fig.~\ref{f2}. To span the
characteristic minimuma in $\Delta\sigma$-versus-$B$ curves, the
fitting interval has been chosen as $-0.3\, B_{tr}<B<0.3\,
B_{tr}$. Strictly speaking, the boundaries of this interval do not
satisfy the diffusion approximation $B\ll B_{tr}$ in which
framework the formulae for magnetoconductance\cite{HLN,Iord} has
been derived. Nevertheless, one can see that the taking into
account only the linear term does not allow us to describe
satisfactorily the magnetoconductivity shape within the fitting
interval while the HLN expression gives a good agreement. Beyond
the diffusion regime, the HLN theory was generalized by Zduniak
{\em et al}.\cite{Zdun} However, the final expressions are very
complicated and inconvenient to use in the fitting procedure.
Because of this, we used the simulation approach described in our
paper, Ref.~\onlinecite{ourtraek}. To take into account the
spin-relaxation processes, Eq.~(20) from this paper was modified
as described in Appendix \ref{app} (details will be published
elsewhere). As Fig.~\ref{f2} illustrates, with the use of
Eq.~(\ref{eq:apRb}) we obtain a nice coincidence of calculated and
measured curves over the whole magnetic field range. Although the
theory beyond the diffusion approximation describes the
magnetoresistance curve better, the fitting parameters
$\tau_1/\tau_\phi$ and $\tau_1/\tau_s$ are found to be close to
those obtained with the help of Eq.~(\ref{eq1}) (see Table
\ref{tab2}). Therefore, it seems natural to analyze the
experimental results with the use of the more simple HLN
expression.

\begin{figure}[t]
\includegraphics[width=\linewidth,clip=true]{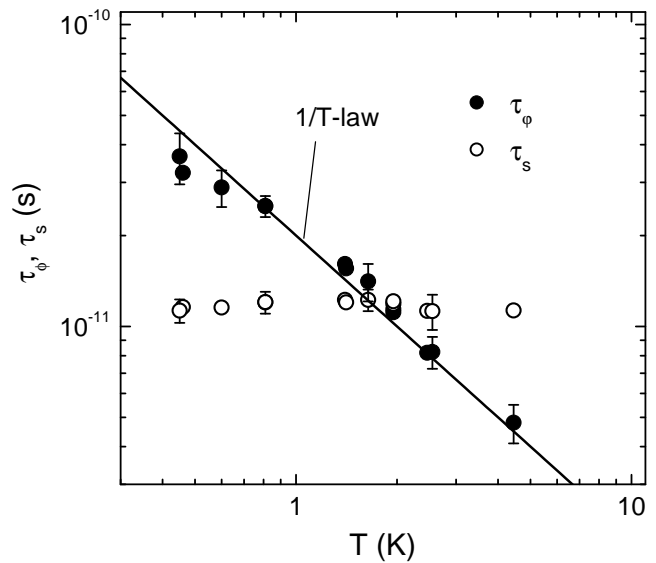}
\caption{\label{f3} The temperature dependence of the phase and
spin relaxation time as obtained from the fit of the experimental
data by Eq.~(\ref{eq1}) for structure 3857 at $V_g=1.5$~V. Solid
line is the $T^{-1}$-law.  }
\end{figure}

\begin{figure}
\includegraphics[width=\linewidth,clip=true]{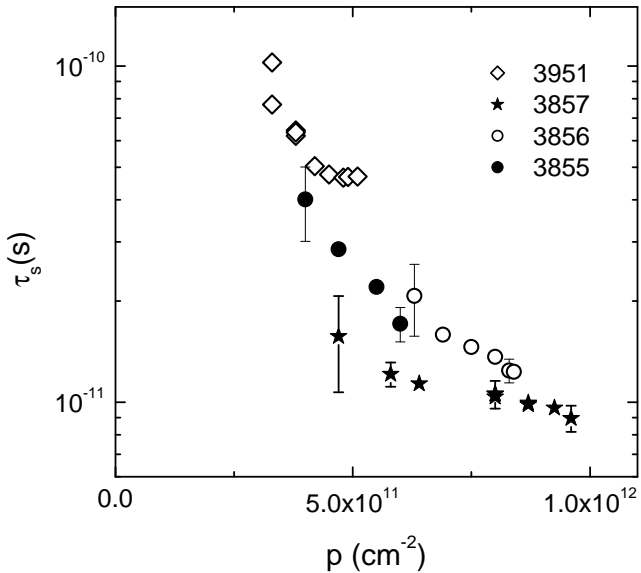}
\caption{\label{f4} The spin relaxation time as a function of hole
density controlled by the gate voltage for all structures
investigated, $T=0.44$~K.}
\end{figure}

Further indication, that just the cubic in $k$ splitting is
responsible for the spin relaxation, is reasonable behavior of the
fitting parameters obtained from Eq.~(\ref{eq1}) with the
temperature change. As seen from Fig.\,\ref{f3} the parameter
$\tau_\phi$ exhibits the behavior close to $T^{-1}$-law that
corresponds to the phase relaxation caused by inelasticity of
electron-electron interaction.\cite{AAK82} The parameter $\tau_s$
is temperature independent as should be for degenerated electron
gas. Such analysis has been carried out for all the structures
investigated and the results are collected in Fig.\,\ref{f4} as a
plot of the spin relaxation time $\tau_s$ against the  hole
density controlled by the gate voltage.

For the first sight the fact that the magnetoconductance curves
are well described by the HLN expression  means that the
Dresselhaus cubic term gives the main contribution to the spin
splitting in the structure investigated. Whether or not it is so,
one can understand analyzing the hole density dependence of
spin-orbital splitting, $\hbar\Omega_3(p)$. For the Dresselhaus
mechanism, the splitting  should be proportional to $p^{3/2}$
because $\Omega_3=\gamma k^3/ 4$, where $\gamma$ is constant
depending only on the band parameters of the parent material (see
Appendix A in Ref.~\onlinecite{Knap96} for details).
Experimentally, the value of spin-orbit splitting $\hbar\Omega_3$
can be found from Eq.~(\ref{eq2}) using $\tau_s$ obtained above
and $\tau_3$. How the quantity $\tau_3$ has been obtained is
considered below.

\begin{figure}[t]
\includegraphics[width=\linewidth,clip=true]{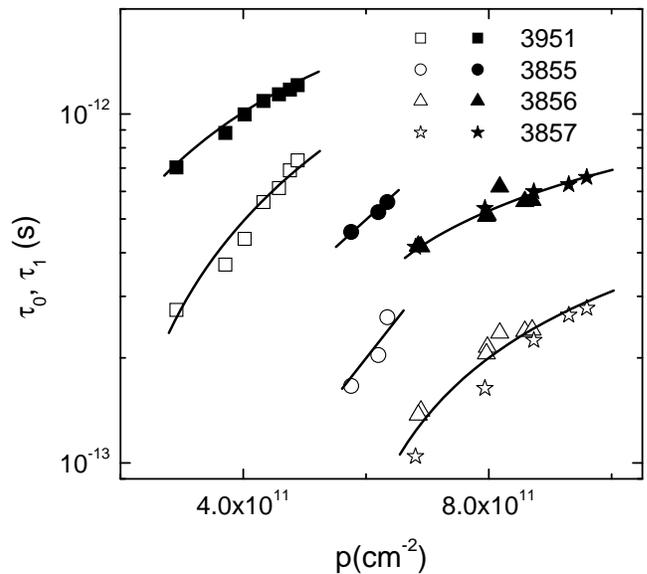}
\caption{\label{f5} The hole density  dependence of $\tau_0$ (open
symbols) and  $\tau_1$ (solid symbols). Solid lines are provided
as a guide for the eye.   }
\end{figure}

As seen from Eq.\,(\ref{eq21}) the relaxation time $\tau_3$ is
determined by the scattering anisotropy via the function
$W(\theta)$. Just the same function determines the relationship
between the quantum and transport relaxation times, $\tau_0$ and
$\tau_1$, respectively. Therefore, we estimate $\tau_3$ using the
experimental value for $\tau_0$
\begin{equation}
\tau_3=\tau_0  \int W(\theta)d\theta \Big/\int
W(\theta)(1-\cos3\theta)d\theta, \label{eq5}
\end{equation}
and conceiving the physically reasonable angle dependence for
$W(\theta)$ so that the ratio
\begin{equation}
K_{01}=\int W(\theta)(1-\cos\theta)d\theta \Big/\int
W(\theta)d\theta, \label{eq6}
\end{equation}
to be equal to the experimental quantity  $\tau_0/\tau_1$. The
value of $\tau_0$ has been obtained from the analysis of the
magnetic field dependence of the amplitude of the Shubnikov-de
Haas oscillations, while $\tau_1$ has been found from the mobility
value, $\tau_1=\mu m/e$. The experimental hole density dependences
of $\tau_0$ and $\tau_1$, presented in Fig.\,\ref{f5}, show that
the scattering is really anisotropic in all the structures  and
the $\tau_0$ to $\tau_1$ ratio lies in the interval from $0.2$ to
$0.5$. This seemingly points to the fact that the scattering is
mainly determined by ionized impurities and $W(\theta)$ can be
chosen in the form obtained, e.g., in Ref.\,\onlinecite{mobInQW}.
However, our estimation shows that the electron mobility in this
case should be one-two order of magnitude higher than that
observed experimentally. We suppose that the roughness of the
quantum well interfaces restricts the mobility in our structures.
This mechanism is theoretically studied in Ref.~\onlinecite{IFR},
where the explicit form for $W(\theta)$ is derived and it is shown
that the scattering anisotropy strongly depends on the parameter
$\Lambda$ characterizing the  fluctuations of the quantum well
width due to interfaces roughness. Using the form for $W(\theta)$
from this paper we have chosen such a value of parameter $\Lambda$
which satisfies the equality between the experimental value of
$\tau_0/\tau_1$ and the calculated from Eq.~(\ref{eq6}) value of
$K_{01}$. Then, with this $\Lambda$-value we have calculated
$\tau_3$ to $\tau_0$ ratio. Doing so we have found that
$\tau_3=(0.7\ldots 0.8)\tau_0$ when $K_{01}$ lies within actual
for our case range, $K_{01}=0.2\ldots 0.5$. Note, the close
results, $\tau_3\simeq\tau_0$, are obtained if one uses
$W(\theta)$ corresponding to scattering by remote ionized
impurities.\cite{mobInQW}

\begin{figure}[t]
\includegraphics[width=\linewidth,clip=true]{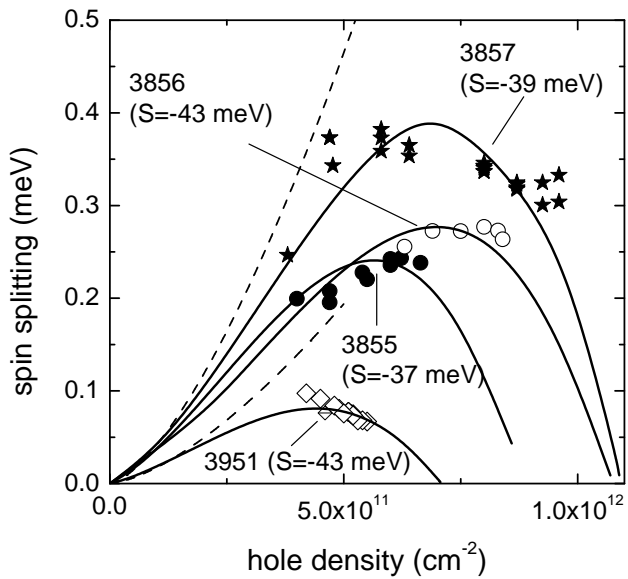}
\caption{\label{f6} The hole density  dependence of the
spin-orbital splitting for different samples. Symbols are the
experimental data obtained as
$\hbar\Omega_3=\hbar/\sqrt{2\tau_3\tau_s}$, solid lines are
calculation results (see text), dashed lines show $p^{3/2}$-law
for structures 3856 and 3951. In brackets,  the values of the
fitting parameter $S$ for each structure are shown.}
\end{figure}

Finally, we arrive the key figure of the paper, Fig.\,\ref{f6},
where the value of spin splitting $\hbar
\Omega_3=\hbar/\sqrt{2\tau_3\tau_s}$ is plotted as a function of
the hole density. One can see that (i) we do not observe the
characteristic for the cubic in $k$ spin-orbit splitting
$p^{3/2}$-dependence and (ii) different structures demonstrate
significantly different value of the splitting for a given hole
density. Both these  facts unambiguously show that no the
Dresselhaus mechanism is responsible for the cubic in $k$
spin-orbit splitting of the hole spectrum.

Let us now discuss  specific features of the Rashba effect for
holes in strained quantum well heterostructures. In general, this
effect in hole 2D systems is considered in
Refs.\,\onlinecite{Gerchikov,Winkler,Winkler1,Xu}. Below we write
out only the main expressions which help us to describe the
experimental results presented  quantitatively. We restrict our
consideration by the case when only three hole bands are taken
into account. They are the heavy- and light-hole $\Gamma_8$ bands
and split off by spin-orbit interaction $\Gamma_7$ hole band. In
this case, the energy spectrum is described by the $6\times6$
Luttinger-Kohn Hamiltonian\cite{LKH} which includes the terms
responsible for the strain.\cite{BP} As shown in
Ref.~\onlinecite{LKH3}, the $6\times6$ Hamiltonian can be
decoupled into two independent $3\times3$ matrices of the form

\begin{equation}
\label{Hmln} H= \left(
    \begin{array}{lll}
    A_+  & C\mp iB           & \sqrt{2} \pm iB/\sqrt{2} \\
    C\pm iB & A_-               & F\mp i \sqrt{3/2}B\\
    \sqrt{2} \mp iB/\sqrt{2}    &F\pm i \sqrt{3/2}B & D
    \end{array}
    \right)
\end{equation}
where
\begin{eqnarray}
A_\pm&=&-(\gamma_1\mp 2\gamma_2)k_z^2-(\gamma_1\pm
\gamma_2)k^2+E_{\Gamma_8}(z)+V(z)\pm S \nonumber \\
B&=&2\sqrt{3}\gamma_3k k_z \nonumber\\
C&=&\sqrt{3}k^2 (\gamma_2^2\cos^2 2\theta + \gamma_3^2\sin^2 2\theta)^{1/2}\nonumber\\
D&=&-\gamma_1(k_z^2+k^2)+E_{\Gamma_7}(z)+V(z) \nonumber \\
F&=&2\gamma_2(\sqrt{2}k^2_z-k^2/\sqrt{2}). \nonumber
\end{eqnarray}
Here, $\gamma_i$ stand for $\hbar^2\gamma_i^L/(2m_0)$, where
$\gamma_i^L$ are the Luttinger parameters, $k_z$ is the wave
vector along the [001] growth direction, $k^2=k_x^2+k_y^2$,
$\theta$ is the angle between the in-plane wave vector and the
[100] direction, $V(z)$ is the macroscopic electric potential in
the heterostructure, $E_{\Gamma_8}$ and $E_{\Gamma_7}$ are the
energies of edges of corresponding bands, and
\begin{equation}
\label{SplitS} S=b\left({\sigma+1\over \sigma-1}\right){\Delta a
\over a}
\end{equation}
is the splitting of the $\Gamma_8$ band due to strain caused by
the lattice mismatch between GaAs and In$_x$Ga$_{1-x}$As. In
Eq.~(\ref{SplitS}), $b$ stands for the axial deformation potential
of the valence band, $\sigma$ is the Poisson's ratio, $\Delta a$
is the lattice mismatch between materials of the quantum well and
barrier, and $a$ is the lattice constant of the quantum well
material. Let us estimate characteristic energies for the case of
GaAs/In$_{0.2}$Ga$_{0.8}$As heterostructure. The value of $\Delta
a/a$ is about 1.4~\%, $\sigma$ is approximately equal to $1/3$,
$b$ is about $-1.7$~eV so that the value of strain induced
splitting $2|S|$ is approximately equal to $90$~meV. This value is
five-ten times greater than the Fermi energy in our case. We find
the Rashba splitting of the hole energy spectrum using the ratio
$E_F/(2S)$ as a small parameter. The band parameters $\gamma_i$
and $\Delta=E_{\Gamma_7}-E_{\Gamma_8}$ are supposed independent of
$z$-coordinate. Then, in isotropic approximation,
$\gamma_2=\gamma_3=\gamma$, the energy spectrum of the upper split
off band for our case can be written as follows
\begin{equation}
E^\pm \simeq E \pm \hbar\Omega_3
\end{equation}
with
\begin{eqnarray}
\hbar\Omega_3 &=& 6\gamma^2 k^3 \int dz\,
|\psi|^2 {d \over dz}\left[{1 \over E-E_{\Gamma_8}(z)-S-V(z)}\right. \nonumber \\
\label{eq:SpSpl} &-&\left.{1 \over
E-E_{\Gamma_8}(z)-\Delta-V(z)}\right],
\end{eqnarray}
where $\psi$ and $E$ are solutions of the  Schr\"{o}dinger
equation
\begin{equation}
A_+\psi=E\psi. \label{eq:Schr}
\end{equation}
It is clearly seen from Eq.~(\ref{eq:SpSpl}), that the Rashba
splitting for all 2D subbands formed from the upper hole band is
cubic in $k$ in contrast to the electron energy spectrum where it
is linear in $k$.

Now we are in position to compare the experimental
$\hbar\Omega_3$-versus-$p$ dependences with those calculated from
Eqs.~(\ref{eq:SpSpl}), (\ref{eq:Schr}). To find the electric
potential $V(z)$, the Schr\"{o}dinger equation has been
self-consistently solved with the Poisson equation. We have used
the parameters $\gamma_i^L$ for In$_{0.2}$Ga$_{0.8}$As and the
value of band offset $\delta
E_v=E_{\Gamma_8}(\text{GaAs})-E_{\Gamma_8}(\text{In}_{0.2}\text{
Ga}_{0.8}\text{As})$, which are obtained by the linear
interpolation from the values of GaAs and InAs: $\gamma_1^L=-9.5$,
$\gamma^L=-3.5$, $\delta E_v=75$ meV, $\Delta=0.35$~eV (the signs
of the Luttinger parameters correspond to the increasing of energy
into the valence band).  As an example, the energy profile and the
wave function for structure 3857, $V_g=0$, is shown in
Fig.~\ref{f0}. To describe the experimental $\Omega_3$-versus-$p$
dependence for each structure, the parameter $S$ has been used as
a fitting one. One can see from Fig.~\ref{f6} that we are able to
describe well the experimental results obtained for different
samples in our  model only slightly varying the $S$-value from the
one to other structure. The different value of strain induced
splitting for different samples seems to be natural.  It can
result, for instance, from deviation of In-content from its
nominal value. As for the value of the strain induced splitting,
$2|S|\simeq 75-90$~meV, it corresponds to the lattice mismatch and
In-content laying within the intervals $(1.2-1.4)$\% and
$(17-20)$\%, respectively. Let us direct attention to the
interesting detail. The $\hbar\Omega_3$-versus-$p$ dependence
exhibits the corresponding to Eq.~(\ref{eq:SpSpl}) behavior,
$\hbar\Omega_3\propto p^{3/2}$, only at low hole density,
$p<2\times 10^{11}$~cm$^{-2}$. At higher hole density this
dependence has a maximum and sign change (not shown in figure).
This feature is caused by the fact that the hole density is varied
by means of variation of the gate voltage. Applying the gate
voltage we change not only the value of the Fermi quasimomentum
but the energy profile of the quantum well as well. In this case
the integral in Eq.~(\ref{eq:SpSpl}) is not constant any more and
gives additional $p$-dependence in $\Omega_3$. Vanishing of
spin-orbit splitting at some hole density means that the quantum
well in this point becomes effectively symmetric. We realize that
the approximations of large strain induced splitting and
$z$-independence of $\gamma_i$ parameters made above are crude
enough. Moreover, the well boundaries can be smooth and different,
and the In-content can vary across the quantum well. These factors
being taken into account could in principle change the value of
$S$ obtained from the fit. However, this should not change our
interpretation in the large.

In summary, we have  shown that the Rashba mechanism results in
the cubic in $k$ spin-orbit splitting of the hole energy spectrum
in  strained heterostructures. The magnetoresistance curve in this
case is well described by the HLN-expression, that allows us to
find the spin splitting as a function of the hole density. We have
found that these dependence is nonmonotonic at relatively high
hole density due to sensitivity of quantum well profile to the
gate voltage.

This work was supported in part by the RFBR (Grants 01-02-16441,
03-02-16150 and 04-02-16626), the CRDF (Grants EK-005-X1 and
Y1-P-05-11), the INTAS (Grant 1B290) and the Russian Program {\it
Physics of Solid State Nanostructures}.

\appendix
\section{Numerical simulation of antilocalization}
\label{app}

The weak localization phenomenon for spin-less particle using the
numerical simulation of a particle motion over the plane with
randomly distributed scatterers is studied both within and beyond
diffusion regime in Ref.~\onlinecite{ourtraek}. It has been shown
that obtaining from the simulation procedure the parameters of
closed paths, one can calculate the quantum interference
correction to the conductivity and its magnetic field dependence
(see Eq.~(20) in the paper cited). Taking into account the spin
relaxation processes leads to the following expression for the
interference quantum correction (in more detail this
generalization will be considered elsewhere)\cite{Igor}
\begin{eqnarray}
\delta\sigma(b)&=&-\frac{2\pi G_0}{I_s d}\sum_{i} \cos( bS_i) \exp
\left(- l_i \gamma \right)
\nonumber \\
& \times & \left[-{1 \over 2}+\exp{\left(-l_i \gamma_s\right)}+{1
\over 2}\exp{\left(-2l_i \gamma_s\right)}\right], \label{eq:apRb}
\end{eqnarray}
where $I_s$ is the total number of paths, $d$ is the diameter of
the area from which the particle starts to walk and in which it
returns,  $l_i$ and $S_i$ are the length and algebraic aria of
$i$-th closed path, $\gamma$ and $\gamma_s$ are the
phenomenological parameters describing the phase and spin
relaxation and corresponding in real systems to ratios
$\tau_1/\tau_\phi$ and $\tau_1/\tau_s$, respectively, the lengths
and areas in this expression are measured in units of mean free
path and squared mean free path, respectively, and summation runs
over all closed paths. In order to treat the experimental results
presented in this paper, we have firstly collected the parameters
of closed paths $l_i$ and $S_i$ simulating the motion of particle
as described in Ref.~\onlinecite{ourtraek}, and,  then, we have
used Eq.~(\ref{eq:apRb}) to fit the experimental curves with
$\gamma$ and $\gamma_s$ as the fitting parameters.

\end{document}